\documentclass{article}
\usepackage{spconf,amsmath,amsfonts,graphicx}

\usepackage{color}
\usepackage{caption}
\usepackage{subcaption}
\usepackage{siunitx}
\usepackage{url}

\usepackage{fancyhdr}

\fancypagestyle{specialfooter}{%
  \fancyhf{}
  
  \fancyfoot[L]{\footnotesize \textcopyright 2020 IEEE. Personal use of this material is permitted. Permission from IEEE must be obtained for all other uses, in any current or future media, including reprinting/republishing this material for advertising or promotional purposes, creating new collective works, for resale or redistribution to servers or lists, or reuse of any copyrighted component of this work in other works.}
}

\newcommand{\tuneenv}{\emph{SoundFilter}}

\begin{document}
\thispagestyle{specialfooter}

\title{One-shot conditional audio filtering of arbitrary sounds}
\name{Beat Gfeller, Dominik Roblek and Marco Tagliasacchi}
\address{Google Research}

\ninept
\maketitle
\begin{abstract}

We consider the problem of separating a particular sound source from a single-channel mixture, based on
only a short sample of the target source. Using \tuneenv, a wave-to-wave neural network architecture, we can train a model without using any sound class labels.  
Using a conditioning encoder model which is learned jointly with the source separation network, the trained model can be ``configured'' to filter arbitrary sound sources, even ones that it has not seen during training.
Evaluated on the FSD50k dataset, our model obtains an SI-SDR improvement of 9.6 dB for mixtures of two sounds.
When trained on Librispeech, our model achieves an SI-SDR improvement of 14.0 dB when separating one voice from a mixture of two speakers.
Moreover, we show that the representation learned by the conditioning encoder clusters acoustically similar sounds together in the embedding space, even though it is trained without using any labels.

\end{abstract}

\section{Introduction}
\label{sec:intro}

A wide variety of sounds exist in the world. In everyday life, people are often exposed to a mixture of various sounds, 
such as birds singing, cars passing by, people talking, and so on.
In this work, we consider the task of taking a mixture of different sound sources, as monophonic audio, and extracting from it a specific sound source. There are various ways in which this task could be approached.
Given the recent success of deep learning for audio processing~\cite{kavalerov2019universal,slizovskaia2020conditioned,ephrat2018looking,li2020learning}, we propose to train a neural network for this task.
A key aspect of our proposal is not to tie the sound to be extracted to any predefined collection of sound categories (such as, for example, the ontology defined by AudioSet~\cite{audioset}).
Instead, we treat the task as a \emph{one-shot learning problem}. The model receives as input the audio mixture to be filtered, together with only one short example of the kind of sound to be filtered. Once trained, the model is expected to extract this specific kind of sound from the mixture if present, and to produce silence otherwise.
We show that this is indeed possible, and propose the following key contributions:
\begin{itemize}
    \item We propose \tuneenv, a wave-to-wave neural network architecture that can be trained end-to-end using 
    single-source audio samples, without requiring any class labels
    that denote the type of source.
    \item We evaluate our method on the
     CC0 subset of the FSD50k~\cite{fsd50k} dataset and the
    Librispeech~\cite{librispeech} dataset, showing a mean SI-SDR improvement of 9.6 dB and 14.0 dB, respectively.
    \item We show that the learned representation encodes semantically meaningful information, even though it is trained without any labels. Specifically, the embeddings of acoustically similar sounds tend to form homogeneous clusters. 
\end{itemize}

\section{Related Work}
\label{sec:related_work}

Closely related to this paper is the work by Kong~et.~al.~\cite{kongIcassp2020}, who also train a neural network for conditional source-separation of single-channel audio. This approach uses a classification model trained on AudioSet~\cite{audioset}, which consists of 10s segments with weak class labels. The trained classification model is then used to extract segments of 1.6 seconds %
for which a class presence is detected with high confidence. The source separation network is then trained on mixtures of two segments from different classes, using one of the two segments as the target. 
The output of the classification model on the target segment are fed to the network as a conditioning input.
At inference time they run the classification network once more, to determine the set of classes present in the input mixture. For each such class, a one-hot vector indicating the selected class is then used to
extract the different sources. 
In short, a key difference between Kong et al.~\cite{kongIcassp2020} and our approach is that the former requires labeled data to train the classifier model, whereas our \tuneenv\ operates in a fully unlabeled setup.
In addition, the embedding used in~\cite{kongIcassp2020} is defined in terms of AudioSet's class ontology. Even though this ontology contains 527 classes, this is still a significant limitation: for example, it does not allow
separation of multiple instances of the same class, e.g., to filter out one specific voice from a mix of multiple voices.

Another closely related area of research is \emph{universal sound separation}. Given a mixture of sounds, the task is to output each of the sound sources.
Kavalerov et al.~\cite{kavalerov2019universal} have recently shown that by constructing a suitable dataset of mixtures of sounds, one can obtain SI-SDR (scale-invariant SDR) improvements of up to 10 dB.
These results can be further improved by conditioning the sound separation model on the
embeddings computed by a pre-trained audio classifier, which can also be fine-tuned~\cite{tzinis2019improving}.
More recently, a completely unsupervised approach to universal sound separation was developed~\cite{wisdom2020unsupervised}, which does not require any single-source ground truth data, but instead can be trained using only real-world audio mixtures.
Another interesting approach which avoids the need for single-source ground truth data uses a pre-trained classifier to enforce that each output channel contains only the sound of one (predefined) class~\cite{DBLP:journals/taslp/PishdadianWR20}.

\begin{figure}[t]
\centering
\includegraphics[width=0.49\textwidth]{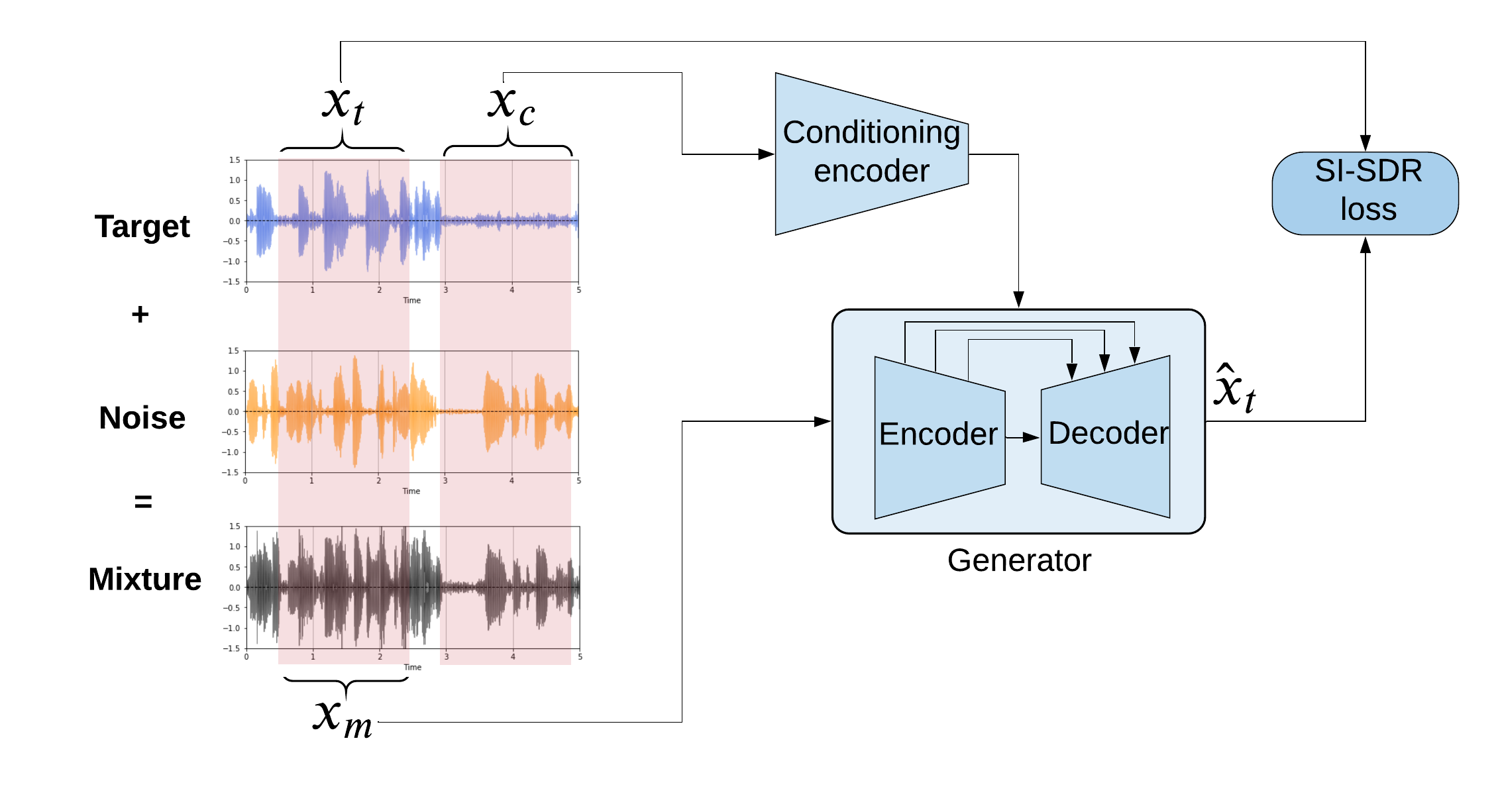}
\caption{\tuneenv~model overview.}
\label{fig:arch-high-level}
\end{figure}

A lot of previous work exists for specific domains, such as
separating speech from either other speech sources~\cite{zeghidour2020wavesplit,voicefilter} or separating speech from other sounds \cite{6853860,DBLP:conf/ica/WeningerEWVRHS15}.
Besides speech, there also has been much work towards extracting individual instrument tracks from music~\cite{stoller2018wave,slizovskaia2019end,slizovskaia2020conditioned}.
Another related line of research performs audio source separation with the help of with other modalities, such as vision~\cite{ephrat2018looking,Xu_2019_ICCV} or accelerometers~\cite{tagliasacchi2020seanet}.

\section{Method}

Our \tuneenv\ model is a wave-to-wave convolutional neural network,  which is trained from mixtures synthesized from a collection of unlabeled audio recordings, in a process outlined below.
Conceptually, we assume that the original audio collection consists of many clips of a few seconds each, which contain the same type of sound for the whole duration of the clip.
We also assume that each such clip contains a single audio source (such as one speaker, one musical instrument, one bird singing, etc.). While not strictly necessary for the training setup to work (see Section~\ref{sec:results}), this simplifies the way that the evaluation operates.
We further assume that the provided original audio data contains a variety of different kinds of sound sources.

\subsection{Training setup}
\label{sec:training}
A training example consists of three parts: i) the target audio, which contains only one sound;
ii) a mixture, which contains two different sounds, one of which is the target audio;
ii) a conditioning audio signal, which is another example containing the same kind of sound as the target audio.
The model is trained to produce the target audio, given the mixture and the conditioning audio as inputs, as illustrated in Figure~\ref{fig:arch-high-level} and detailed below. We use training examples where all three parts have equal length $L$, say 2s, and proceed as follows:
\begin{enumerate}
    \item  Select an example from the training dataset and extract two random crops $x_t \in \mathbb{R}^L$ and $x_c \in \mathbb{R}^L$ of length $L$ each, where $x_t$ denotes the target audio and $x_c$ the conditioning audio.
    \item Select another distinct random example $x_n \in \mathbb{R}^L$ from the training dataset, which is used as noise source.
    \item Create a mixture $x_m \in \mathbb{R}^L$, by mixing $x_t \in \mathbb{R}^L$ (target) with $x_n \in \mathbb{R}^L$ (noise). During training, we mix the two examples at varying SNRs, chosen uniformly at random between $-4$~dB and $+4$~dB, to create mixtures of varying difficulty.
\end{enumerate}

\subsection{Model architecture}

The model architecture takes as input two sequences of audio samples (the mixture audio and the conditioning audio), and outputs the filtered audio $\hat{x}_t$, as illustrated in Figure~\ref{fig:arch-high-level}.
Our model consists of  two components: 
i) a \emph{conditioning encoder}, which takes the conditioning audio and computes the corresponding embedding, and
ii) a conditional \emph{generator}, which takes the mixture audio and the conditioning embedding as input, and produces the filtered output.

\begin{figure}
     \centering
     \begin{subfigure}[b]{0.26\textwidth}
         \centering
        \includegraphics[width=0.98\textwidth]{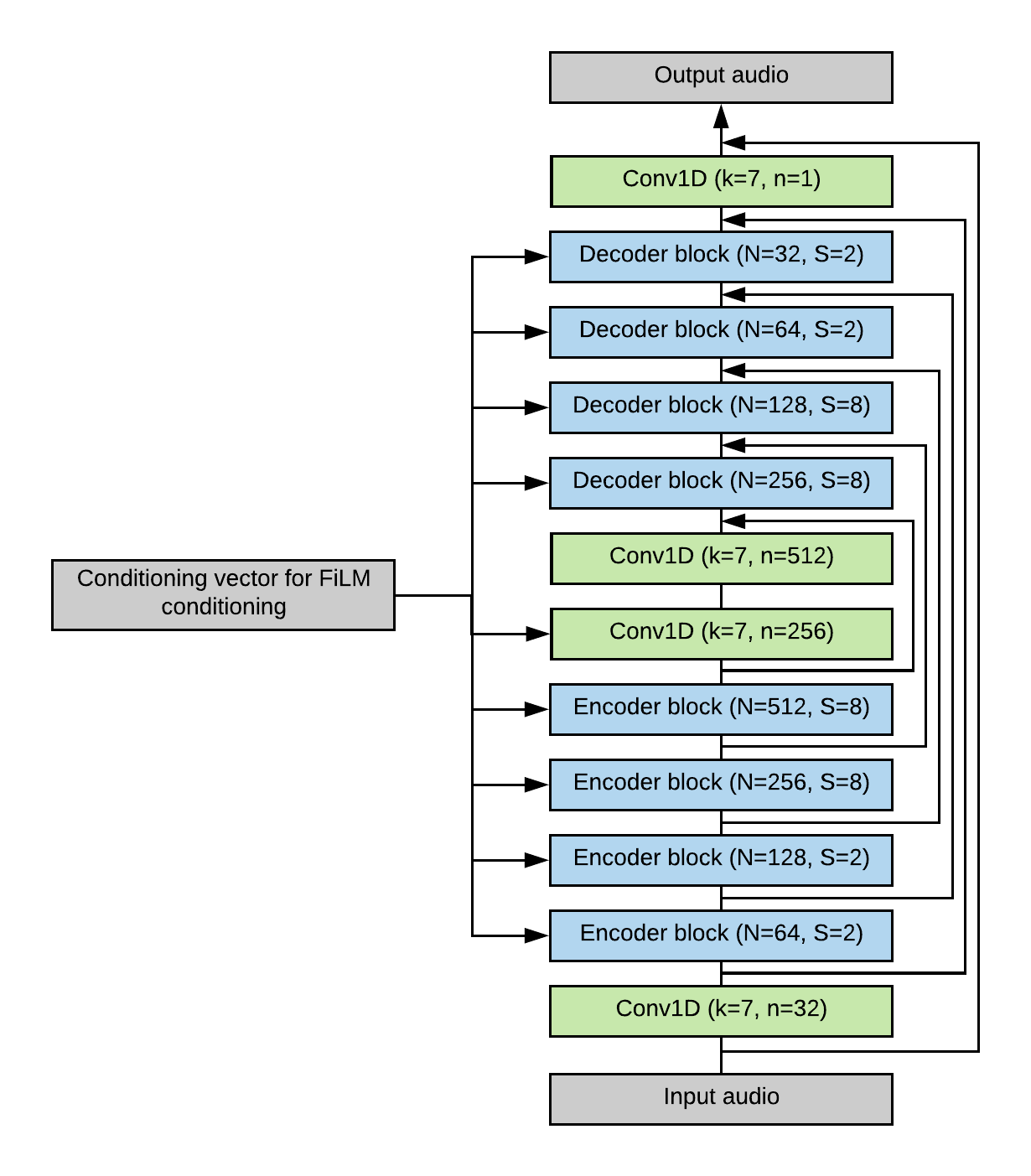}
     \end{subfigure}
     \begin{subfigure}[b]{0.20\textwidth}
         \centering
        \includegraphics[width=0.98\textwidth]{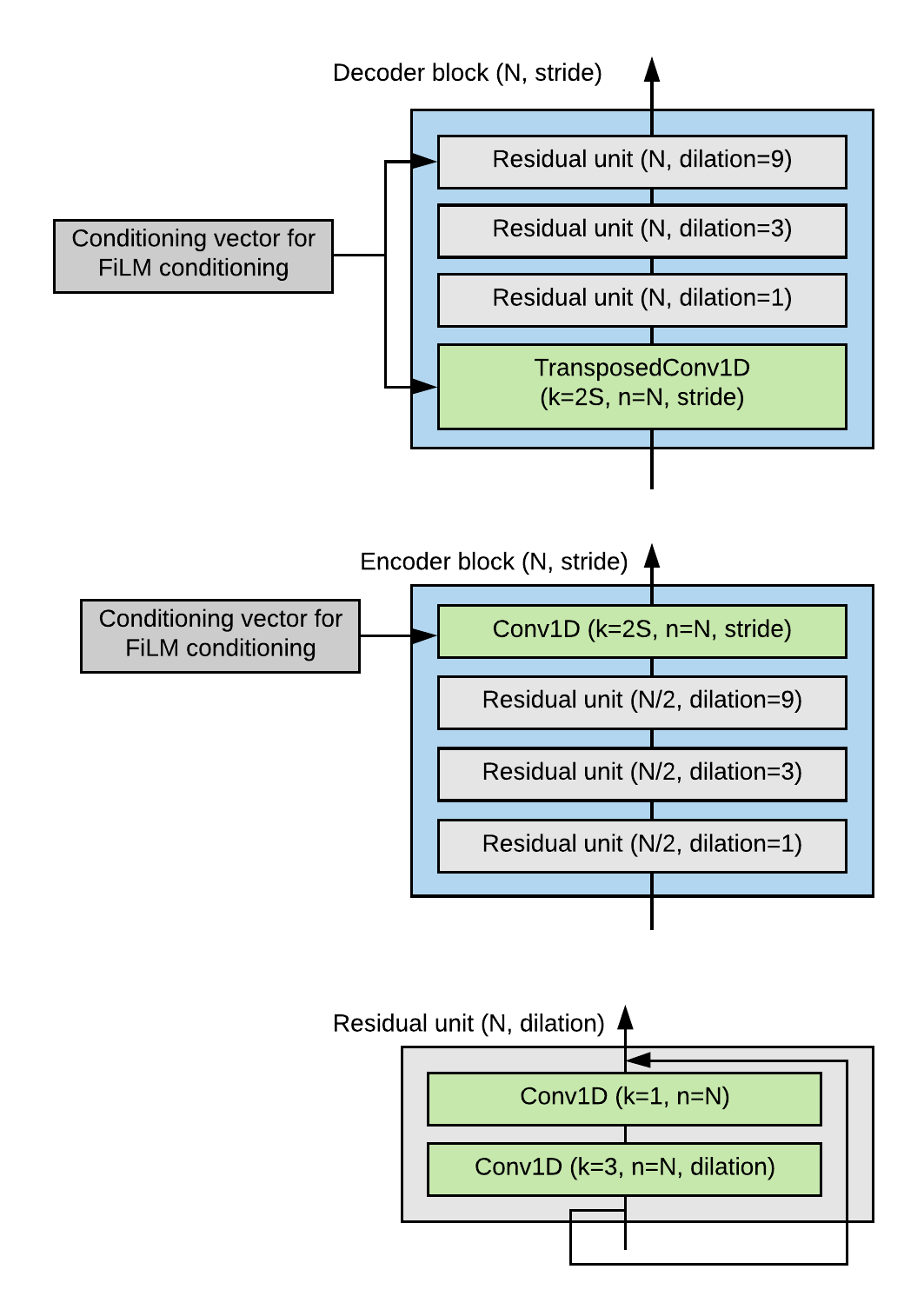}
     \end{subfigure}
        \caption{Detailed outline of the conditioned audio-to-audio U-Net architecture.}
        \label{fig:unet_architecture}
\end{figure}

The generator is a wave-to-wave U-Net~\cite{unet2d} architecture, whose architecture is detailed in Figure~\ref{fig:unet_architecture}. In a nutshell, it is a symmetric
encoder-decoder network with skip-connections, where the architecture of the decoder layers mirrors the structure of the encoder, and the skip-connections run between each encoder block and its mirrored decoder block.
This architecture consists of an encoder, which outputs a bottleneck embedding, followed by
a decoder, which generates the output samples from the bottleneck embedding, the skip-connections and the conditioning input. 
The encoder and the decoder each have
four blocks stacked together, which are sandwiched between
two plain convolution layers. The encoder follows a downsampling scheme of (2, 2, 8, 8) while the decoder up-samples in
the reverse order. The number of channels is doubled whenever
down-sampling and halved whenever up-sampling. Each encoder block consists of an up-sampling layer in form of a transposed 1D convolution, followed by a residual unit made up of
three 1D convolutions with a kernel size of 3 and dilation rates
of 1, 3, and 9, respectively. The decoder block also mirrors the
encoder block, and consists of the same residual unit followed
by a strided 1D convolution for down-sampling. 
A skip-connection
is added between each encoder block and its mirrored decoder
block. The outermost skip connection feeds the input directly to the output of the last convolution, to which the input is added.
We use group normalization after each Conv1D operation (not shown in the figure for simplicity)~\cite{groupnorm}.
Moreover, we use the ELU activation function~\cite{clevert2016fast} for all non-linearities in the model.

A key ingredient of the proposed model is that the generator is conditioned on the embedding representation computed from a sample acoustically similar to the target. Note that while a similar U-Net architecture was used in~\cite{tagliasacchi2020seanet}, in that case the conditioning signal was fed directly to the encoder as an additional audio channel. Instead, in this work we implement the conditioning mechanism using FiLM~\cite{perez2017film,DBLP:conf/nips/BirnbaumKEKE19}, which is inserted after some of the layers of the U-Net architecture, as illustrated in Figure~\ref{fig:unet_architecture}.
For simplicity, we use exactly the same architecture for the conditioning encoder as for the encoder of the U-Net, with the difference that the former has no conditioning input. Due to the fully convolutional architecture, the conditioning encoder produces a sequence of $d$-dimensional embeddings, at a sampling frequency  256 times smaller than the original audio sampling frequency ($2\times2\times8\times8 = 256$). 
Since we assume that the same sound source is present throughout the length of the (relatively short) input mixture $x_m$,
we aggregate this sequence of embeddings along the temporal dimension, into a single $d$-dimensional embedding (we used $d=256$ in our experiments), before feeding it to the conditional generator.
A simple way to aggregate these embeddings in time is to use max-pooling.
In order to better adapt to non-homogeneous conditioning audio, we propose to use a
cosine-similarity based attention mechanism (called \emph{content-addressing} in~\cite{DBLP:journals/corr/GravesWD14}).
Namely, we first compute a sequence of embeddings for the target audio using the conditional encoder and apply max-pooling to obtain a single embedding $v_{t}$. 
Then, we compute the weighted average of the embeddings produced from the conditioning audio,
using as weights the softmax-compressed cosine similarities between $v_t$ and the individual embeddings. L2-normalization is applied to all embeddings before and after aggregation.
In order to use the final embedding in each FiLM conditioning layer of the conditional audio model,
these values are linearly projected to the required number of values (i.e., the number of output channels of the preceding convolution).

The generator and the conditioning encoder are jointly trained end-to-end, with the objective of maximizing the scale-invariant signal-to-distortion ratio (SI-SDR)~\cite{roux2018sdr} (soft-clipped at a limit of 30 dB) between the target~$x_t$ and the filtered audio~$\hat{x_t}$. The SI-SDR measures the SDR between a target and estimate within an arbitrary scale factor, which is
widely used for evaluating source separation tasks.
Our training setup runs on GPU or TPU, is implemented in TensorFlow and uses the Adam optimizer with a learning rate of $1e^{-4}$, default values of $\beta_1=0.9$, $\beta_2=0.999$, and a batch size of $32$.

\begin{figure*}[t]
\centering
\includegraphics[width=0.95\textwidth]{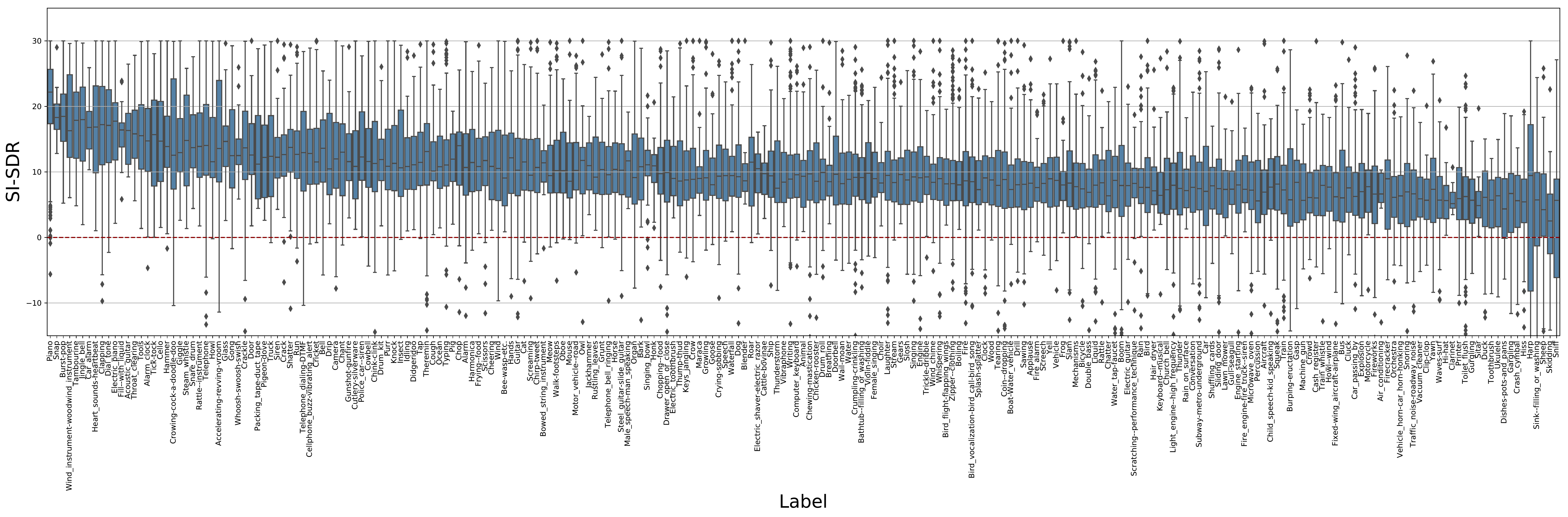}
\caption{SI-SDR metrics for FSD50k per class, sorted by mean. We recommend reading the electronic version, which allows zooming in arbitrarily, to read the class labels.}
\label{fig:sisdr-classes}
\end{figure*}

\section{Experiments}

We evaluate our model on examples that are synthetically generated in a similar way as the training data described in Section~\ref{sec:training}. During evaluation,  we ensure that target audio and the conditioning audio are always \emph{disjoint} crops from the same input example, which never overlap in time.
In contrast to training, where we mix target audio and noise at different SNR levels, during evaluation we always mix with at a SNR equal to 0dB.
As the evaluation metric, we use SI-SDR, as it is invariant to gain mismatches. We report the average and the standard deviation over 5 different runs. 
Note that neither the training nor the evaluation of our model requires the availability of labeled samples, as it merely relies on the assumption that audio examples contain the same sound sources (one or multiple) throughout the example. 

\subsection{Datasets}

The main dataset used for training and evaluation is the recent FSD50k dataset~\cite{fsd50k}\footnote{We only use samples with a CC0 license.}, which contains recordings of various types of sounds provided by users via Freesound, since our focus of interest is the filtering of single sound sources of any kind. 
This dataset comes with manually-verified labels for 200 classes, defined as a subset of the AudioSet ontology~\cite{audioset}, which we use to compute per-class metrics. 
In addition, we evaluate \tuneenv~for the task of separating overlapping speakers, for which we use the Librispeech dataset~\cite{librispeech}, containing short clips of speech from 251 different speakers.

\subsection{Results}
\label{sec:results}

\begin{table}[t]
\centering
\caption{SI-SDR improvement (SI-SDRi) averaged across replicas.}\label{tab:main_results}
\addtolength{\tabcolsep}{0.1cm}
\begin{tabular}{c S[
                 table-number-alignment = center,
                 separate-uncertainty = true,
                 table-figures-uncertainty = 1,
                 table-figures-decimal = 1,
                 table-figures-integer = 1
         ]}
Dataset & \text{SI-SDRi} [dB]  \\         
\hline
FSD50k & 9.6 \pm 0.2 \\
Librispeech & 14.0 \pm 0.1 \\
\hline
\end{tabular}
\end{table}

We trained two separate \tuneenv~models on FSD50k and Librispeech and obtained an average SI-SDR improvement equal to 9.6dB and 14.0dB, respectively, as reported in Table~\ref{tab:main_results}. Using the labels available in FSD50k, we analyzed the results for each class separately, as illustrated in Figure~\ref{fig:sisdr-classes}. We observe that some classes are easier to separate from the mixtures than others. For example, for \emph{Piano}, \emph{Tambourine}, \emph{Car Alarm} and \emph{Clapping}, the mean SI-SDR is well over 15 dB. For almost all classes we achieve a mean SI-SDR  above 5 dB, except for a few classes such as \emph{Crash cymbal}, \emph{Harp} and \emph{Toothbrush}.
We could not identify a clear pattern between the acoustic properties of the sounds and the observed SI-SDR, except that  separating short sounds such as \emph{Snap} seems to be easier. Audio samples produced by \tuneenv\ are publicly available\footnote{\url{https://google-research.github.io/seanet/soundfilter/examples/}}.

To evaluate \tuneenv~in more challenging and realistic \emph{one-shot} conditions, we trained the model on examples for which both the target and the noise audio belong to a set of classes, while we evaluate on examples created from a completely disjoint set of classes. This experiment is meant to evaluate how well the model generalizes to new classes not seen during training. To this end, we used the labels to prepare 5 different splits of the FSD50k dataset, each containing a disjoint subset of 46 classes. 
Then, we trained 5 separate models, where for each model $i=1,2,\ldots, 5$ we omit the classes in group $i$ from the training data, and use only classes from the remaining group $i$ during evaluation. Table~\ref{tab:split-results} shows the results for each different split. We observe that the average SI-SDR varies across splits, due to the intrinsic difficulty of the classes sampled in each split, but it remains above 8.6dB on the worst split, and is on average equal to 9.4dB, i.e., only marginally worse than the SI-SDR achieved when training and evaluating using all classes.

\begin{table}[t]
\centering
\caption{SI-SDR improvement (SI-SDRi) for FSD50k, when training and evaluation is on disjoint classes, averaged across replicas.}\label{tab:split-results}
\addtolength{\tabcolsep}{0.1cm}
\begin{tabular}{c S[
                 table-number-alignment = center,
                 separate-uncertainty = true,
                 table-figures-uncertainty = 1,
                 table-figures-decimal = 1,
                 table-figures-integer = 1
         ]}
Split & \text{SI-SDRi [dB]}  \\         
\hline
 split 1  & 9.4 \pm 0.0  \\
 split 2  & 9.1  \pm 0.2  \\
 split 3  & 8.8  \pm 0.2  \\
 split 4  & 10.0  \pm 0.2  \\
 split 5  & 9.8  \pm 0.2  \\
 \hline
 avg. of all splits & 9.4  \pm 0.4  \\
\hline
\end{tabular}
\end{table}

We also performed an ablation study over different design choices of our model to better understand their impact. Specifically, we repeated the evaluation with the following variants of the proposed model:
\begin{itemize}
    \item \emph{Model capacity}: We varied the model capacity by adjusting the number of channels used in all convolutional layers. Instead of using 32 as the initial channel depth and doubling it in each encoder block (Figure~\ref{fig:unet_architecture}), we scaled all depths by 2x, 4x, and 0.5x. Table~\ref{tab:ablation-results} shows that increasing model capacity in this way does improve SI-SDR, at the cost of a considerable  increase in the number of parameters.
    \item \emph{Attention mechanism}: We observe that removing the attention mechanism based on cosine similarity, and instead aggregating the sequence of embeddings produced by the conditioning encoder by directly using max-pooling along the temporal direction, the SI-SDR remains at 9.6dB. 
    Despite this result, we believe the attention mechanism to be beneficial in practice, in particular for non-homogeneous conditioning inputs.
    \item \emph{Normalization}: We found that the choice of the normalization applied to the model activations is critical for the good performance of the model. Indeed, replacing group normalization (using groups of 16 channels) with batch normalization, the observed SI-SDR decreases to 7.1dB.
    \item \emph{Conditioning encoder}: The conditioning encoder in \tuneenv~is trained end-to-end with the generator. We tried to replace this component with a fixed encoder, which was trained as a fully supervised classifier on Audioset. We use the logit layer of the classification model (which produces one value for each of the 527 Audioset classes) and project it to 256 dimensions with a learnable linear layer.
    \item \emph{Loss function}: We replaced the SI-SDR loss function with a multi-scale spectrogram loss as proposed in~\cite{engel2020ddsp} (computed in the mel-spectrogram domain as suggested in~\cite{DBLP:journals/corr/abs-2008-01160}). Unsurprisingly, the SI-SDR decreases to 8.8dB, since there is a mismatch between the loss function used during training and the evaluation metrics. However, the quality of the output is very similar to the one obtained using the SI-SDR loss and a more comprehensive subjective evaluation campaign is needed to fully evaluate the impact of the choice of the loss function.
    \item \emph{Single-source audio:} Both FSD50k and Librispeech have the property that each example contains only one sound source, present throughout the example. To investigate the impact of dropping this requirement, we also trained \tuneenv\ using AudioSet~\cite{audioset}, which mostly consists of mixtures of multiple sources. Notably, we observe a relatively small drop in performance when evaluating on FSD50k, from 9.6dB to 8.6dB.
\end{itemize}

\begin{table}[t]
\centering
\caption{SI-SDR improvement (SI-SDRi) on FSD50k, for different model variants, averaged across replicas. }\label{tab:ablation-results}
\addtolength{\tabcolsep}{0.1cm}
\begin{tabular}{c S[
                 table-number-alignment = center,
                 separate-uncertainty = true,
                 table-figures-uncertainty = 1,
                 table-figures-decimal = 1,
                 table-figures-integer = 1
         ]}
Split & \text{SI-SDRi [dB]}  \\         
\hline 
Baseline  & 9.6 \pm 0.2  \\
\hline
    Conv depth: 16  & 8.8 \pm 0.3  \\
    Conv depth: 64  & 9.8  \pm 0.1  \\
    Conv depth: 128  & 10.0 \pm 0.1  \\
    Without attention mechanism & 9.6 \pm 0.0 \\
    Batch norm instead of group norm & 6.9 \pm 0.5 \\
    Pre-trained AudioSet embeddings & 7.9 \pm 0.3 \\
    Multi-scale resolution loss (DDSP) & 8.0 \pm 0.2 \\
    Training on AudioSet (eval on FSD50k) & 8.6 \pm 0.2 \\
 \hline    
\end{tabular}
\end{table}

\subsection{Visualizing the learned embedding space}
\begin{figure}[t]
\centering
\includegraphics[width=0.22\textwidth]{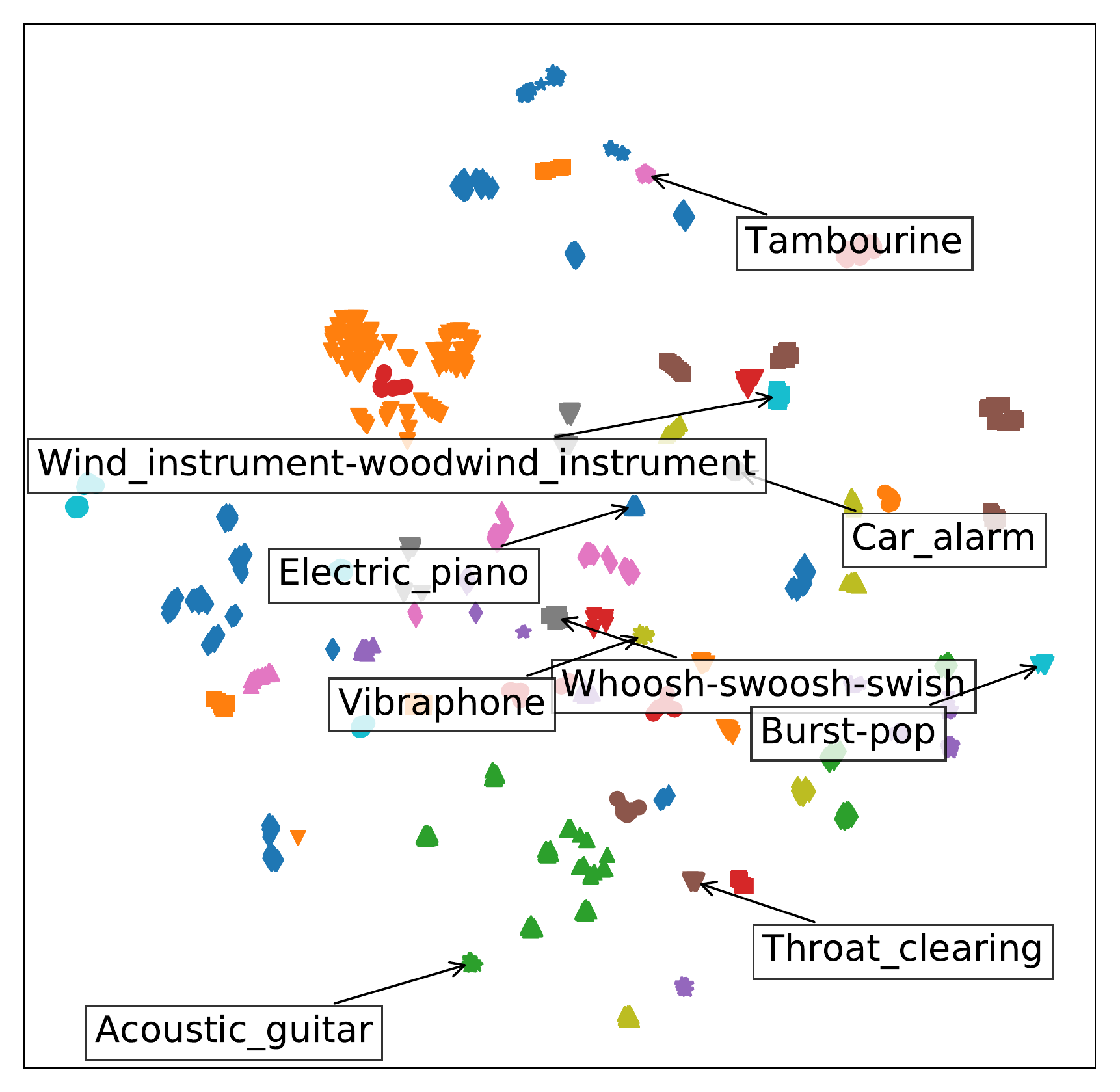}
\includegraphics[width=0.22\textwidth]{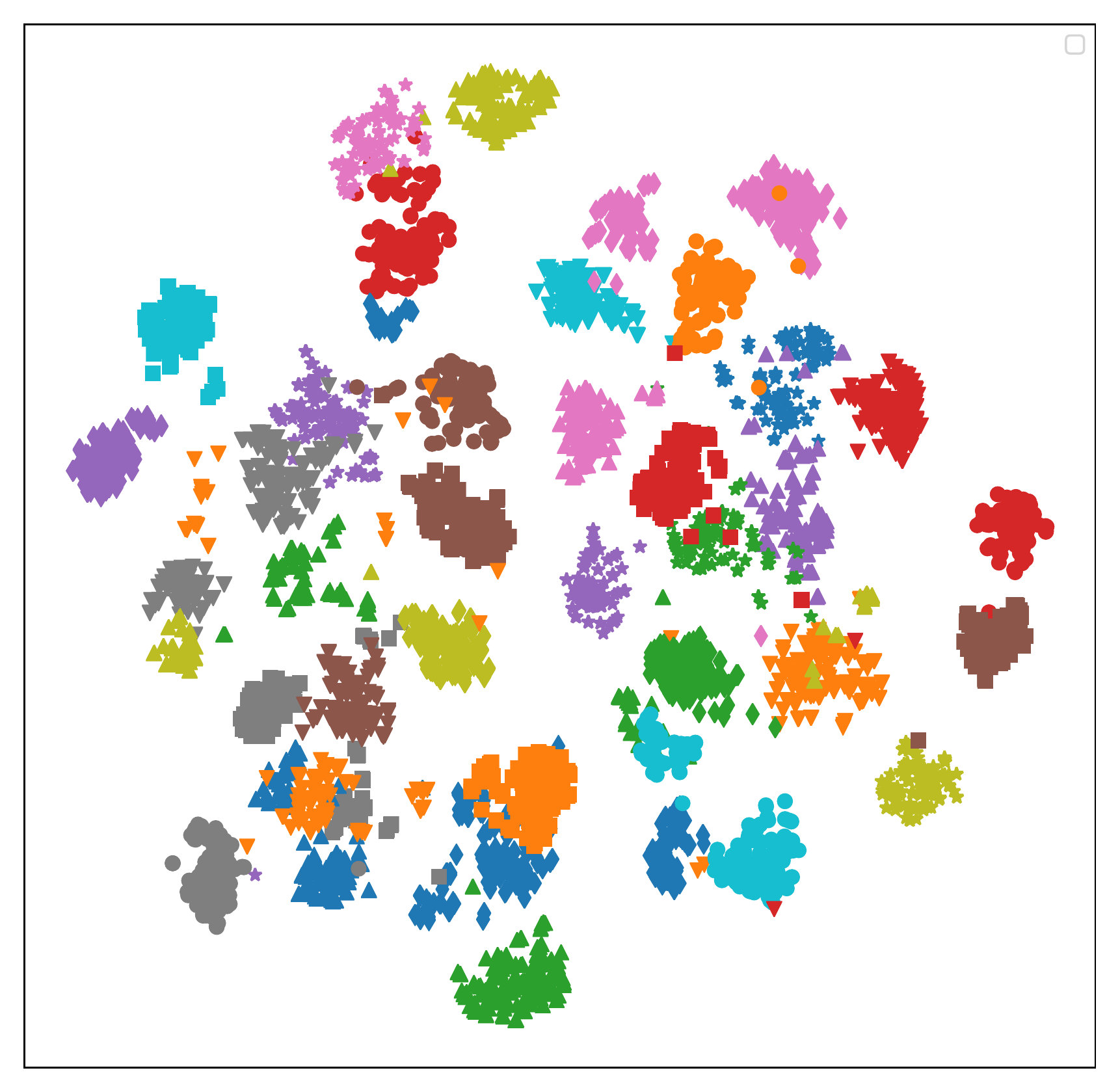}
\caption{t-SNE plots of learned embeddings for FSD50k (left), colored by class label and Librispeech (right), colored by the speaker identity. Both plots are showing embeddings for evaluation data, unseen during training. We recommend reading the electronic version, which allows zooming in arbitrarily, to read the labels.}
\label{fig:tsne-embeddings}
\end{figure}

The proposed \tuneenv~model is trained without having access to class labels. However, the conditioning encoder learns to produce embeddings that represent the acoustic characteristics of the conditioning audio. To verify this, we visualize a 2-dimensional projection of the embedding space with t-SNE~\cite{tsne}, by clustering the embeddings of the examples seen during evaluation. Figure~\ref{fig:tsne-embeddings} shows the learned representation, for both FSD50k and Librispeech, where different colors/markers denote, respectively, different class labels and speakers. In both cases we observe a clear structure, which correlates well with class labels, despite the fact that labels were not used during training. This observation gives some evidence that the conditioning embeddings encode semantically meaningful information.

\section{Conclusion}
We present \tuneenv, a conditional sound separation model that can focus on an arbitrary type of sound in a  mixture of sounds, given only a short sample. It can be trained using only unlabelled examples of single-source recordings.
Our work could be extended by
exploring how to use the embedding learned as part of \tuneenv\ as a representation for an audio event classifier.
In addition,
it would be of interest to extend our approach from one-shot to many-shot.

\clearpage

\bibliographystyle{IEEEbib}
\bibliography{references}
\label{sec:refs}

\end{document}